\documentclass{revtex4}
\usepackage{graphicx}
\usepackage{amssymb}
\usepackage{amsmath}

\begin{document}

\author{Francesco Sorrentino${}^{\ddagger*1}$, Nicholas Mecholsky${}^{*}$ }
\affiliation{${}^\ddagger$ Universit{\`a} degli Studi di Napoli Parthenope, 80143 Napoli, Italy \\${}^*$ IREAP,  University of Maryland, College Park, Maryland 20742\\ ${}^1$ Corresponding author. E-mail: {\tt
fsorrent@unina.it}}

\begin{abstract}
We consider a network of coupled agents playing the Prisoner's Dilemma game, in which players are allowed  to pick  a strategy in the interval $[0,1]$, with $0$ corresponding to defection,  $1$ to cooperation, and intermediate values representing mixed strategies in which each player may act as a cooperator or a defector over a large number of interactions with a certain probability. Our model is payoff-driven, i.e., we assume that the level of accumulated payoff at each node is a relevant parameter in the selection of strategies. Also, we consider that each player chooses his/her strategy in a context of limited information.
We present a deterministic {nonlinear} model for the evolution of strategies. 
{We show that the final
strategies depend on the network structure and on the choice of the parameters of the game.} 
{We find that polarized strategies (pure cooperator/defector states) typically emerge when (i) the network connections are sparse, (ii) the network degree distribution is heterogeneous, (iii) the network is assortative, and surprisingly, (iv) the benefit of cooperation is high.
}
\end{abstract}

\title{Stability of strategies in  payoff-driven evolutionary games on networks}

\maketitle

\textbf{We study a Prisoner's Dilemma game on a network to uncover the effect of the network structure on game dynamics.
We describe a model for an individual's changing strategy based on a payoff comparison with the player's neighbors. This type of model is relevant to many situations where strategies change due to payoffs such as in politics, economics, or finance.}

\section{Introduction}

The Prisoner's Dilemma is a paradigmatic model for interactions among agents where strategies leading to either individual gain (defection) or the common good (cooperation) are in competition.
In politics, economics, and in finance, an individual's actions may be regarded as cooperation or defection.
Sometimes, the choice of a strategy (whether to cooperate or defect) is affected by
direct reciprocity
but other times, the choice of strategy is affected by the performance of players in the game. 



Previous research has focused on how  direct reciprocity in a repeated game (i.e., trust between connected individuals) can affect the evolution of  strategies \cite{DR1,DR2,DR3,DR4,DR5}.  Alternatively, players' performance may affect their goals and future choice of strategies.
For example, direct reciprocity cannot capture altruistic or selfish behavior.
Therefore, a problem of interest is understanding,  when the players are coupled in some complex way over a network, how their strategies evolve in time, based on their performance in the game. This 
is the subject of the current paper. 

Evolutionary game theory studies how players' strategies evolve when the game is iterated in time. Here we consider that the agents are coupled over a complex network and that the game is played between each agent and his/her direct neighbors over the network. 
 An important difference with previous literature is that our analysis is based neither on a population level analysis \footnote{A classical model of this type involves replicator dynamics, which describes the time-evolution of frequencies of players  who adopt given strategies in a large population. In general, such a macroscopic description is appropriate when the underlying social network is characterized by a mean-field type interaction and the number of agents is very large.  The case of finite-size populations has been studied e.g., in \cite{Arne06}. The effects of departure from the well-mixed population assumption on the evolution of cooperation have been studied in \cite{GT2,GT3,GT4,GT5,GT1,GT6}. } nor on a mean field approximation; instead, we are interested in how a specific network structure affects the evolution of strategies, and in particular the fixed points of the dynamics and their stability. For a review on evolutionary games on networks, see \cite{GT_REV}.

In recent years, much research effort has been devoted to studying the structure and dynamics of complex networks (see e.g., \cite{Wa:St98,Ba:Al99,Pa:Ve00a,korea,Ne:Gi02,New02Ass,Ba:Pe02,New03Mix,Ar:dB:So,restr06c,Report,NewSIAM}). However, a full understanding of how the specific network  structure affects the evolution of strategies for models in evolutionary game theory is lacking. Here we present a unified analytical treatment which holds for diverse networks including those with random, scale free, and degree-correlated topologies.

In the literature on game theory, a distinction is commonly made between pure and mixed strategies. In the first case, players are allowed to choose from a finite set of alternatives (e.g., in the case of two strategies, either defection or cooperation). In the second case, each player is allowed to choose  a strategy in a continuum range, e.g., $[0,1]$, with $0$ ($1$) representing pure defection (pure cooperation). In a probabilistic framework, a mixed strategy equal to $0.25$ corresponds to cooperating $1/4$ of the time and defecting $3/4$ of the time. In many real social situations, mixed strategies are common, as rarely an individual is observed to behave as a cooperator (defector) all of the time. It is also reasonable to expect that mixed strategies will sometimes be convenient, as they allow a player to hedge against the risk of choosing a pure strategy (in the same way as diversifying a portfolio reduces the risk of an investment). In what follows, we will consider a network of coupled agents, each of which can choose a strategy in a continuum of values between pure defection and pure cooperation. Agents are then able to update their individual strategy  based on  the interactions with the neighbors. Pure strategies are eventually recovered as particular cases of mixed strategies (i.e, $0$ or $1$). Hence, the pure strategies lie on the boundary of the mixed strategies.  As a result of this, we will see that pure strategies play a particular role (qualitatively different) in the dynamics. 

In what follows, we introduce a model for the evolution of strategies that we analyze. The main features of our model are the following: I) it is payoff driven; II) each agent has limited information on the other players' strategies. Below we describe points I) and II) in more detail. 

I)  Most of the ongoing research on evolutionary game theory focuses on situations where the agents are equal, i.e., they respond in the same way when presented with the same stimulus. However, this assumption is unrealistic. In this paper, we model a situation where each node's choice of strategy depends on an internal parameter that may differ from node to node. We want this parameter to evolve with time based on the interactions with the other players.   We consider that each individual's choice of strategy depends on his/her degree of success in playing the game; specifically, we consider that the level of accumulated payoff at each node becomes a relevant parameter.
{As motivation, we note how, in social contexts, the behaviors of individuals are highly correlated with their social status; also, in economics {\emph{powerful agents/players}} are usually more competitive than weaker ones.   Maslow's theory of motivation \cite{Maslow} ranks human needs in a hierarchy (usually represented as a pyramid) and argues that as each lower need is satisfied, the next higher level becomes more compelling, i.e., there is a shift in the perception of the relevant needs at each level of the hierarchy.}
In what follows, we will incorporate this feature in our model of an iterated game, 
i.e., once the accumulated payoff of player $i$ exceeds a threshold, this will have an effect on the choice of strategy of $i$.  

II) Most of the current research in evolutionary game theory considers that individuals choose their strategy based on knowledge of other players' strategies. However, this assumption is
unrealistic. In this paper, we consider a situation in which knowledge about other players' strategies may be unavailable, while we assume that each player has knowledge
of his/her neighbors' payoffs. These assumptions apply to those situations in which the strategies that  players adopt are  hidden or undeclared. For example, we may expect that in certain situations a defector may be reluctant to openly declare his/her strategy.
Note that if information on the payoffs is available, this can provide indirect evidence of the strategies adopted by the players; i.e., for example, if a neighbor is driving a Ferrari, this might indicate that he or she must be a strong cooperator or a strong defector; but which one would be the right guess?    \footnote{Answering this question is nontrivial, as we are putting ourselves in a typical situation characterized by limited information. In this paper we focus on situations in which agents have to make decisions in the absence of information on their neighbors' strategies. Therefore, this kind of questions represents a motivation for our study. Moreover, we note that the fact that there is not an obvious answer to this question, makes it an interesting subject of investigation.
}


To conclude, our model presents both these characteristics: it is payoff-driven and it does not presume knowledge of the other players' strategies. As such, it aims at encompassing certain sources of complexity that characterize real-world situations.  Given our model, we are able to carry out a stability analysis that is highly dependent on the specific network structure and provides a unified framework to study the effects of different network features, such as random, scale free, and degree-correlated topologies.

{In Sec.\ II we present our model for the evolution of strategies of coupled agents over a network. In Sec.\  III we classify the fixed points of the dynamics in fully mixed strategies and polarized strategies and we present a stability analysis that gives a simple condition on the eigenvalues of a relevant matrix. The role of the underlying network topology on stability is considered in Sec.\ IV. Conclusions are presented in Sec.\ V.}

\section{Model Formulation}

We consider a network of coupled agents (nodes) playing the Prisoner's Dilemma game.  Each node $i=1,...,N$ is characterized by a strategy $\sigma_i$ in the closed interval $[0,1]$, where the strategy $\sigma_i$ represents the probability that player $i$ behaves as a cooperator over a large number of interactions ($1-\sigma_i$ is the probability that player $i$ behaves as a defector). The  case of $\sigma_i=0$ ($\sigma_i=1$) corresponds to  a \emph{pure strategy}, in which player $i$ is always a defector (respectively, a cooperator). We refer to all the remaining cases, i.e., $\sigma_i \in(0,1)$, as \emph{mixed strategies}. We assume that  the game is played very frequently so  that at our resolution we are unable to assess whether agent $i$ behaves as a cooperator or a defector at any given interaction, but only the average number of times that it acts as a cooperator, i.e., the frequency $\sigma_i$, which we regard as the \emph{state} of node $i$. Hence we assume to only have access to the probability $\sigma_i$ (while information about whether a node behaves as a cooperator or defector at any given interaction remains unavailable), and we regard $\sigma_i$ as the \emph{state} of node $i$. 
Over many interactions, individuals may occupy any position in the range of probabilistic defector/cooperator behavior when considering their record of behavior.

We consider that the evolution of the strategies is governed by the nodes' payoffs, where the  payoff of node $i$ is determined by its strategy and those strategies of the nodes that are connected to it. In the Prisoner's Dilemma (see e.g., Ref. \cite{GT2,GT4}),
a cooperator incurs a cost $c$ for each connection that he or she has. If a cooperator is connected to another cooperator, the benefit of that connection is $b$. So, if a cooperator is connected to $n$ individuals and $m$ of them are cooperators, the payoff for that cooperator is $b m - c n$. Typically, $b>c$. For a defector, there is no cost associated with any connection, and if the defector is connected to $l$ cooperators, the defector's payoff is $bl$.

We assume that the network structure could be directed and weighted and is described by the adjacency matrix $A=\{A_{ij}\}$, where $A_{ij}>0$ if $j$ is connected to $i$, $0$ otherwise.
This yields the following definition of the payoff for a player $i$,
\begin{equation}
p_i=\sigma_i[b \sum_j A_{ij} \sigma_j-c \sum_j A_{ij}]+(1-\sigma_i)b \sum_j A_{ij} \sigma_j=b \sum_j A_{ij}\sigma_j-c  \sum_j A_{ij} \sigma_i=\sum_j A_{ij} (b \sigma_j -c \sigma_i), \label{pay}
\end{equation}
or in matrix notation,
\begin{equation}
\bar{p}=(bA-cD){\bar \sigma}, \label{payvec}
\end{equation}
where $\bar{p}=[p_1,p_2,...,p_N]$ is the vector of payoffs, $\bar{\sigma}=[\sigma_1,\sigma_2,...,\sigma_N]$ is the vector of strategies, and $D=\{d_{ij}\}$ is a diagonal matrix, such that $d_{ii}=\sum_j A_{ij}$.
Again, $p_i$ in (\ref{pay}) is the average payoff of node $i$ over a large number of interactions with its neighbors, when its strategy and those of the neighbors are  given by fixed  $\sigma$'s.

We allow the strategies adopted by each node/player to evolve in time;  thus we write  $\sigma_i=\sigma_i(t)$, where $t$ represents continuous time. Our basic assumption is that each node updates its strategy based on its payoff and the payoffs of its neighbors (note we do not make the assumption that nodes know the strategies of their neighbors. Indeed, that knowledge could be unavailable). With this information, player $i$ can compute the relative payoff,
\begin{equation}
u_i=\sum_j A_{ij} [p_i(t)-p_j(t)], \label{u}
\end{equation}
with a positive (negative) value of $u_i$ indicating that node $i$ is performing better (poorly) with respect to its neighbors.

From (\ref{payvec},\ref{u}), we obtain,
\begin{equation}
\bar{u}=(D-A)\bar{p}=B \bar{\sigma}, \label{central}
\end{equation}
where $\bar{u}=[u_1,u_2,...,u_N]$, and the matrix $B$ is defined as,
\begin{equation}
B=(D-A)(bA-cD).
\end{equation}

From (\ref{central}) we now see that the relative payoff of node $i$ can  be written as a linear combination of the strategies, i.e., $u_i=\sum_j B_{ij} \sigma_j$, $i=1,...,N$. Note that the matrix $B$ encodes information of both the network structure (i.e., the matrix $A$) and the choice of the parameters of the game (i.e., the two scalars $b$ and $c$). We will show that the eigenvalues of this matrix control stability of a fixed set of strategies. 

We aim at formulating a simple, general model for the evolution of the strategies $\{\sigma_i(t)\}$, evolving from $\sigma_i(0) \in (0,1)$, that fulfils the following requirements,

  (i) The pure strategies $\sigma_i=\{0,1\}$ are fixed points for the dynamics, as $\sigma_i(t)$ cannot decrease below $0$ or exceed $1$.

  (ii) The dynamics admits other fixed points for values of $\sigma_i \in (0,1)$, corresponding to mixed strategies in the spectrum of possibilities between complete cooperation and complete defection.

  (iii) The stability of the fixed points [(i),(ii)] depends on an {internal} adaptive parameter  at each node {(which we label $\mu_i$, $i=1,..,N$)} whose evolution  is governed by the interactions with the neighboring nodes. Under specific conditions, $\mu_i$ is a measure of the \emph{wealth} of $i$, i.e., the accumulated \emph{payoff} at node $i$. We assume that when $\mu_i$ exceeds a threshold, this will affect the choice of strategy of $i$. 



With these conditions in mind, we write
the following set of differential equations,
\begin{subequations}\label{d}
\begin{align}
\dot{\sigma}_i(t)= & \alpha \sigma_i(t) (\sigma_i(t)-1) (\sigma_i(t)-\mu_i(t)), \label{de}\\ \label{de2}
\dot{\mu}_i(t)= & f(u_i(t)),
\end{align}
\end{subequations}
$i=1,...,N$.  Equation (\ref{de}) determines the evolution of the strategies, where $\alpha>0$, $\mu_i$ is the adaptive parameter, whose evolution is specified by Eq.\ (\ref{de2}).  Here, $\mu_i$ is the $i^{\textrm{th}}$ node's internal parameter that models the player's changing attitude toward his neighbors.  The evolution of $\mu_i(t)$ is forced by the input
$u_i(t)=\sum_j B_{ij} \sigma_j(t)$,
through the coupling function $f$ 
that we require be continuous and strictly monotonically increasing/decreasing with $f(0)=0$. For $f(u)=u$, $\mu_i$ is simply the integral of $u_i(t)$, the relative payoff of node $i$, over time, and as such it represents the \emph{relative wealth} of node $i$
(see (iii) above). Here, again, the word \emph{relative}, means with respect to node $i$'s neighbors.
{We wish to emphasize that
from Eq. (\ref{d}), our model is payoff-driven (it evolves based on a comparison
of
the
payoff between node $i$ and its neighbors) and does not presume knowledge at node $i$ of its neighbors' strategies $\sigma_j$, $j \neq i$.}

The parameter $\mu_i$ determines the stability of the fixed points of Eq.\ (\ref{de}). The differential equation (\ref{de}) has three fixed points: $\sigma_i=\{0,1,\mu_i\}$. The form of (\ref{de}) is such that for any initial condition $\sigma_i(0) \in (0,1)$, the trajectories are constrained to lie in the closed interval $[0,1]$. We now evaluate stability of the fixed points of (\ref{de}), when the values of $\mu_i$ are fixed. This is summarized in Fig.\ 1.
For $\mu_i \in (0,1)$, the pure strategies $\sigma_i=\{0,1\}$ are unstable, while the mixed strategy $\sigma_i=\mu_i$ is stable.
For $\mu_i \geq 1$ ($\mu_i \leq 0$), the fixed point $1$ is stable and the fixed point $0$ is unstable (the fixed point $1$ is unstable and the fixed point $0$ is stable).
\begin{figure}[t]
\centering
\includegraphics[width=0.8\textwidth]{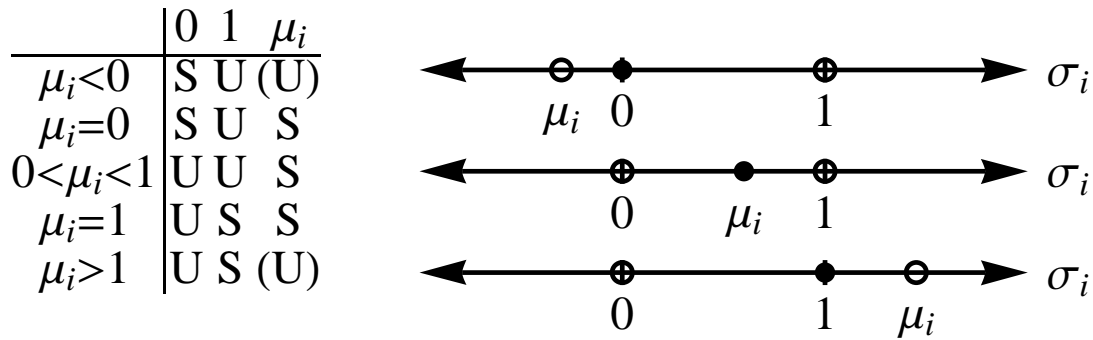}
\caption{\small The table (on the left) shows the stability (S) or instability (U) of each of the fixed points $\{0, 1, \mu_i \}$ for any occurrence of $\mu_i$. Note that the stability of the fixed points $\sigma_i=\mu_i$ for $\mu_i>1$ and for $\mu_i<0$ is in brackets as these equilibria are never observed if $\sigma_i(0) \in (0,1)$. On the right, we have a graphical representation of the stability of the fixed points of Eq.\ (\ref{de}) for different values of $\mu_i$, where open circles represent unstable fixed points and filled circles represent stable ones. From top to bottom, we show the cases of $\mu_i<0$, $0<\mu_i<1$, and $\mu>1$.}
\label{A0}
\end{figure}

When $\mu_i$ increases (decreases) above 1 (below 0), it affects the stability of the fixed points of Eq. (\ref{de}) as shown in Fig.\ 1. Specifically, for $f(u)=u$, when $\mu_i$ exceeds (goes below) a threshold, this results in a qualitative change of the behavior of $i$.  This corresponds to a transition from one level to another of the Maslow's pyramid of needs, resulting in a change of the individual's attitude (and, as a consequence, his/her strategy). 

{In our numerical simulations to follow}, we choose  the following form for the function $f$,
\begin{equation}
f(u)=\beta \tanh(\gamma u), \label{f}
\end{equation}
$\beta,\gamma>0$. With this choice, $\mu_i$ can still be seen as the wealth accumulated at node $i$ but with a saturation on the maximum increment or decrement that is allowed. {There are many possible situations in which gains and losses are typically bounded.
As an example, consider a taxation system that charges a tax on capital gains, while capital losses are tax-deductible.}

The form of the function $f$ in (\ref{f}) is consistent with the observation that favorable (unfavorable) conditions often foster cooperative (defective) behavior \cite{SVO2,SVO3,SVO4} (e.g., empirical studies have found a positive correlation between the level of cooperation and the payoff for players participating in a Prisoner's Dilemma experiment \cite{SVO2}).  However, {we wish to emphasize that} the results that we present in this paper ({and in the particular, the stability analysis of Sec. III}) are, to a great extent, independent of the specific choice of the function $f$. {On another note, we observe that the function $f$ describes how the relative payoff influences the strategy adoption, hence it defines an \emph{imitation function} as defined in Ref. \cite{Binwu}.
}

\section{Stability Analysis}

In this section we consider the stability of the fixed points of Eqs.\ (\ref{d}), when the parameters $\mu_i$ are not fixed (i.e., the individual systems $i=1,...,N$ are coupled).
We first look for fixed points $\{\sigma_i^*\}_{i=1}^{N}$ of the dynamics for Eqs.\ (\ref{d}). 
A candidate fixed point $\{\sigma_i\}_{i=1}^{N}$ may include both pure and mixed strategies for different nodes, i.e., for some $i$'s $\sigma_i = \{0,1\}$, and for the remaining $i$'s, $\sigma_i=\mu_i \in (0,1)$.
We refer to a fixed point for which $\sigma_i^* \in (0,1)$, for all $i=1,...,N$ as a \emph{fully mixed strategy}, otherwise, we refer to it as a \emph{polarized strategy}. 

In what follows, without loss of generality, we assume that for any equilibrium  point, the indices $i$ are labeled so that $\sigma^*_i = \{0,1\}$, $i=1,...,\ell$, $0 \leq \ell \leq N$,  and $\sigma^*_i \in (0,1)$, $i=\ell+1,...,N$. 
With this relabeling, we note that the  equations (\ref{de}), $i=1,...,\ell$,  are decoupled from the corresponding equations (\ref{de2}), $i=1,...,\ell$. It follows that the first $\ell$ equations (\ref{d}) drive the last $(N-\ell)$ equations (\ref{d}).
 Thus we define any set of strategies to be a fixed point $\{\sigma_i^*\}_{i=1}^{N}$ if the following conditions are met: (I) Eqs.\ (\ref{de}) are set to zero  for $i=1,...,\ell$ and (II)  Eqs.\ (\ref{de},\ref{de2}) are simultaneously set to zero for $i=\ell+1,...,N$ ($\mu_i^*=\sigma_i^*$, $i=\ell+1,...,N$).

{Given a vector of strategies, $\{\sigma_i\}_{i=1}^N$, we would like to derive conditions on this vector that identify it as a fixed point. Conditions (I) are always satisfied if
$\sigma_i \in \{0,1\}$ for all $i=1,...,\ell$.}  Conditions (II) become,
$\sum_{j=\ell+1}^N B_{ij} \sigma_j= -\sum_{j=1}^{\ell} B_{ij} \sigma_j$, $ i=\ell+1,...,N$,
or in matrix form,
\begin{equation}
B^r \bar{\sigma}^r= \bar{c}, \label{sis}
\end{equation}
which is a system of ($N-\ell$) linear equations in ($N-\ell$) unknowns; the matrix $B^r$ is a reduced matrix, which is obtained by eliminating the first $\ell$ rows and the first $\ell$ columns from the matrix $B$, $\bar{\sigma}^r$ is an $(N-\ell)$-vector which is obtained by eliminating the first $\ell$ entries from the vector $\bar \sigma$, and $\bar{c}=[c_1,c_2,...,c_{N-\ell}]$ is an $(N-\ell)$-vector, which is determined by the first $\ell$ entries of $\bar \sigma$, i.e., $c_i=-\sum_{j=1}^{\ell} B_{ij} \sigma_j$. If $\bar{c}=\bar{0}$, the system of linear equations (\ref{sis}) is homogeneous (case (A)), so any $\bar{\sigma}^r$ in the null subspace $\textrm{Ker}(B^r)$ of the matrix $B^r$ is a solution. 
Otherwise, (\ref{sis}) is an inhomogeneous system (case (B)) and it admits one and only one solution, provided that $B^r$ is invertible, i.e., $\bar{\sigma}^r= (B^r)^{-1} \bar{c}$. Fully mixed strategies (for which $\ell=0$, $B^r \equiv B$) are included in case (A)\footnote{In Appendix A we show that the matrix $B$ has at least one zero eigenvalue. Therefore there is at least one eigenvector in the null subspace of the matrix $B$, other than the null vector $\bar{0}$.}.

We consider $\sigma_i(0) \in (0,1)$, which constrains $\sigma_i(t)$ to be in the interval $[0,1]$. {A vector in  $\textrm{Ker}(B^r)$ does not necessarily have to represent a set of physical strategies. For example, for a given matrix $B^r$, it is possible that there are vectors in $\textrm{Ker}(B^r)$ that have both positive and negative entries. Then, even with the multiplicative constant that we are allowed for eigenvectors, it might be impossible to rescale all entries of such vectors 
into the range $[0,1]$.} Thus, in order for $\bar{\sigma}^r$ to be a physically meaningful equilibrium for the set of equations (\ref{de}), $i=\ell+1,...,N$, it has to satisfy the additional requirement (III) that: for case (A), all the entries of $\bar{\sigma}^r$ have the same sign (i.e., they are all positive or all negative) and, for case (B),  the entries of the vector ${\bar \sigma}^r$ are in the interval $(0,1)$. {For the remainder of the paper, we assume the dimension of $\textrm{Ker}(B^r)$ to be $1$; in which case, the single zero eigenvector $\bar{v}_0$ of $B^r$ is the basis for $\textrm{Ker}(B^r)$. Then any vector $\bar{\sigma}^r= a \bar{v}_0$, where $a \in \mathbb{R}$ is a fixed point but not necessarily physical.}

We see that the system of equations (\ref{d}) behaves as a multistable system for which the final attractor depends on the choice of the initial conditions. 
Here, we include a stability analysis for the fixed points of (\ref{d}), i.e., for points $\{\sigma_i^*\}_{i=1}^N$ that satisfy the above conditions (I),(II), and (III). We already know that the first $\ell$ equations (\ref{de}) are decoupled from the others and admit only one stable equilibrium (if $0$ is stable, $1$ is unstable, and viceversa). Therefore, in order to study stability of an equilibrium $\{\sigma_i^*\}_{i=1}^N$, we need to linearize Eqs.\ (\ref{de2}), for $i=\ell+1,...,N$, about $\{\sigma_i^*\}_{i=1}^N$ (with $\sigma_i^*=\mu_i^* \in (0,1), i=\ell+1,...,N$),
\begin{subequations} \label{delta} \begin{align}
\delta \dot{\sigma}_i= & \alpha ({\sigma_i^*}^2-\sigma_i^*) (\delta \sigma_i-\delta \mu_i), \label{deltaa}\\
\delta \dot{\mu}_i= &  Df(0) \sum_{j=\ell+1}^N B_{ij} \delta \sigma_j,\label{deltab}
\end{align}\end{subequations}
$i=\ell+1,...,N$, where, since $0<\sigma_i^*<1$, we have that $-0.25<({\sigma_i^*}^2-\sigma_i^*)<0$.  We now write $\sum_{j=\ell+1}^N B_{ij} \delta \sigma_j=\lambda_k \delta \sigma_i$, $i=\ell+1,...,N$,  where $\lambda_k$  is an eigenvalue and [$\delta \sigma_{\ell+1},...,\delta \sigma_{N}$] is an associated eigenvector for the matrix $B^r$, $k=1,...,(N-\ell)$. By substituting in (\ref{deltab}) we obtain,
\begin{subequations} \label{deltak} \begin{align}
\delta \dot{\sigma}_i= & \alpha ({\sigma_i^*}^2-\sigma_i^*) (\delta \sigma_i-\delta \mu_i), \label{deltaak}\\
\delta \dot{\mu}_i= &  Df(0) \lambda_k \delta \sigma_i,\label{deltabk}
\end{align}\end{subequations}
$i=\ell+1,...,N$, and $k=1,...,(N-\ell)$. Note that Eqs.\ (\ref{deltak}) represent a set of $(N-\ell)^2$ equations, each of which is independent of the others. Equations (\ref{deltak}) can be rewritten, 
\begin{equation}\label{systemi}
\left[
     \begin{array}{c}
       \delta {\dot {\sigma}}_i \\
       \delta {\dot {\mu}}_i \\
     \end{array}
   \right]= \left[
     \begin{array}{cc}
       \alpha \phi_i & -\alpha \phi_i \\
       Df(0) \lambda_k & 0 \\
     \end{array}
   \right]\left[
     \begin{array}{c}
       \delta {\sigma}_i \\
       \delta {\mu}_i \\
     \end{array}
   \right],
\end{equation}
$i=\ell+1,...,N$, $k=1,...,(N-\ell)$, where $\phi_i=({\sigma_i^*}^2-\sigma_i^*)<0$ and $\lambda_k=\lambda_k^r+j \lambda_k^i$, where $j^2=-1$

We first proceed under the assumption that the $\lambda_k$'s are real (which is the case e.g., when the matrix $A$ is symmetric, i.e., the network is undirected, see Appendix B). We see from (\ref{systemi}),  that stability is independent of $\phi_i$, as long as $\phi_i$ is negative, but depends on the conditions,
\begin{equation}
Df(0) \lambda_k \leq 0,  \quad k=1,...,(N-\ell). \label{ldr_bis}
\end{equation}
The case of $\lambda_k=0$ corresponds to $\delta {\tilde \sigma}$ decaying to zero, with $\delta {\tilde \mu}$ constant. {Hence, this $\lambda_k=0$ eigenvalue corresponds to neutral stability along the direction of the eigenvector $\bar{v}_0$.}  For $Df(0)>0$ ($Df(0)<0$), condition (\ref{ldr_bis}) is equivalent to $\lambda_{max}\leq0$ ($\lambda_{min}\geq0$),  where $\lambda_{max}=\max_{k=1,...,(N-\ell)} \lambda_k$ ($\lambda_{min}=\min_{k=1,...,(N-\ell)})$.
Note that this is a necessary and sufficient condition for stability.

In the case the eigenvalues $\lambda_k$ of the matrix $B^r$ are complex, stability requires that the real parts of both the eigenvalues of the system (\ref{systemi}) be negative for $i=\ell+1,...,N$ and $k=1,...,(N-\ell)$. This corresponds to the following conditions,
\begin{subequations}
\begin{align}
Df(0) \lambda_k^r & \leq 0, \quad \\
\quad |Df(0) \lambda_k^i| & < \sqrt{\alpha \phi_i Df(0) \lambda_k^r },
\end{align}
\end{subequations}
which need to be satisfied for $i=\ell+1,...,N$ and for $k=1,...,(N-\ell)$. In what follows, we proceed under the assumption that the eigenvalues $\lambda_k$ are real.


One nice property of the stability condition (\ref{ldr_bis}) is that it  decouples the effect of the  coupling function $f$ from the effects of the eigenvalues $\lambda_k$ of the matrix $B^r$ (of the matrix $B$ in the case $\ell=0$), where the eigenvalues $\lambda_k$ reflect the particular choice of the network structure (i.e., the matrix $A$) and the form of the game (i.e., the parameters $b$ and $c$).
Also, condition (\ref{ldr_bis}) allows us to predict whether some network/game combinations would support fully mixed strategies or promote polarized behaviors. {For example, given the network in Fig.\ 2(d), we are able to determine that for $b=1.1$ there is a stable fully mixed strategy and for $b=1.5$ it is unstable.}


\begin{figure}[t]
\centering
\includegraphics[width=0.5\textwidth]{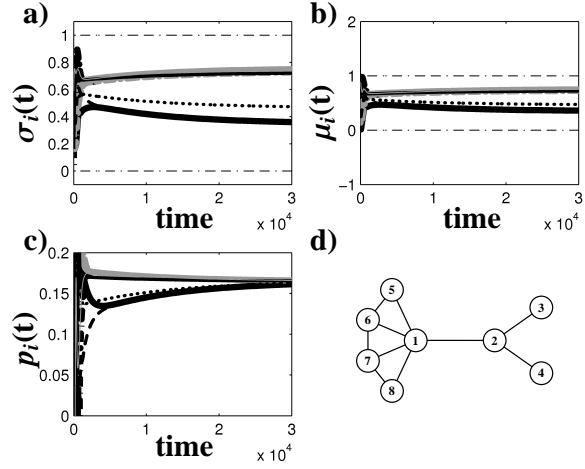}
\caption{\small  We consider an $8$-node network, shown in plot d. The benefit of cooperation is set to $b=1.1$ ($c=1$).
Plot a shows the time evolution of $\sigma_i(t)$, $i=1,\ldots,8$, with all the strategies  ${\sigma_{i}(t)}$ converging on a fully mixed state, corresponding to $\bar{v}_0$. Plot b shows the evolution of $\mu_i(t)$, $i=1,\ldots,8$. Plot c shows the evolution of $p_i(t)$, $i=1,\ldots,8$, with all the payoffs converging after a transient to $0.165$.
The initial conditions for $\mu$'s and $\sigma$'s are randomly chosen from a uniform distribution in the interval $(0,1)$, $\alpha=5 \times 10^{-2}$, $\beta=10^{-3}$, $\gamma=1$.}
\label{A1a}
\end{figure}

\begin{figure}[t]
\centering
\includegraphics[width=0.5\textwidth]{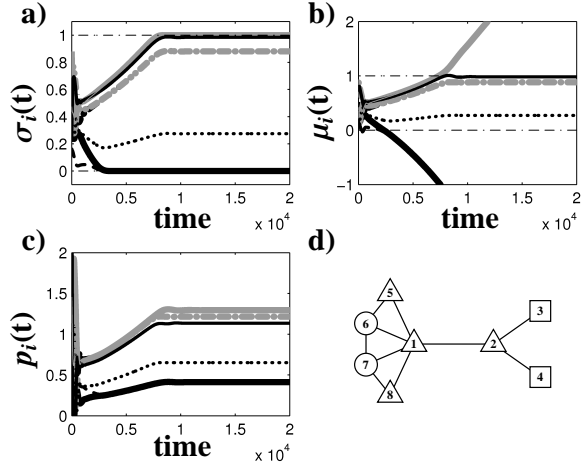}
\caption{\small We repeat the simulation in Fig.\ 2, but now the benefit of cooperation $b=1.5$.   With this choice of the parameters, the fully mixed state is unstable. Plot a shows the time evolution of $\sigma_i(t)$, $i=1,\ldots,8$. Nodes 3 and 4 (shown as squares in the graph) converge on the defector state (0); nodes 6 and 7 (shown as circles in the graph) converge on the cooperator state (1); the remaining four nodes (shown as triangles) converge on a mixed strategy. Plot b shows the evolution of $\mu_i(t)$, $i=1,\ldots,8$. Plot c shows the evolution of the payoffs $p_i(t)$, $i=1,\ldots,8$, which converge on different final values: $p_1=1.13, p_2=0.652,p_3=0.411,p_4=0.411,p_5=1.21,p_6=1.29,p_7=1.29,p_8=1.21$.
The initial conditions for $\mu$'s and $\sigma$'s are randomly chosen from a uniform distribution in the interval $(0,1)$, $\alpha=5 \times 10^{-2}$, $\beta=10^{-3}$, $\gamma=1$.}
\label{A1b}
\end{figure}

Figures \ref{A1a} and \ref{A1b}  show the effect of the choice of the parameters of the game on the stability of the strategies. Fig.\ \ref{A1a} (\ref{A1b}) shows a simulations for $b=1.1$ ($b=1.5$), where $b$ is the benefit of being connected to a cooperator ($c=1$  in both cases). For $b=1.1$, the eigenvalues of the matrix $B$ are $\{-38.7,  -20.2,  -12.4,   -9.15,   -2.42,   -1.00,   -0.0841,   0\}$, i.e., they are all negative. The eigenvector associated with the eigenvalue $0$ is $\bar{v}_0=[0.405,    0.260,    0.194,    0.194,    0.409,    0.423,    0.423,    0.409]$, i.e., its components are all of the same sign. Thus we expect a fully mixed strategy proportional to $\bar{v}_0$, to be a stable fixed point for the dynamics. The initial conditions $\sigma_i(0)$ and $\mu_i(0)$ are randomly selected  from a uniform distribution in the interval $(0,1)$. Fig.\ \ref{A1a}(a) (\ref{A1a}(b)) shows the evolution of the individual $\sigma_i(t)$ (of the $\mu_i(t)$) versus $t$, with the $\sigma_i$ (and the $\mu_i$) converging on a solution proportional to $\bar{v}_0$. Fig.\ \ref{A1a}(c) shows the time evolutions of the payoffs $p_i(t)$, $i=1,...,N$, which after a transient converge on the same value $p_i=0.165$, $i=1,...,N$.

For $b=1.5$, the eigenvalues of the matrix $B$ are $\{ -42.9,  -22.9,  -13.5,   -9.74,   -2.06,   -1.00,    0,    0.148\}$, and $\bar{v}_0=[0.203,    0.562,    0.556,    0.556,    0.0706,    0.0824,    0.0824,    0.0706]$. Since one of the eigenvalues is positive, we do not expect to see a solution proportional to $\bar{v}_0$. The case of $b=1.5$ is shown in Fig.\ \ref{A1b}. As can be seen from Fig.\ \ref{A1b}(a), after a transient, nodes 3 and 4 (shown as squares in the graph in Fig.\ \ref{A1b}(d)) converge on the defector state $0$; nodes 6 and 7 converge on the cooperator state $1$ (shown as circles). The remaining four nodes converge on a mixed strategy (shown as triangles). The eigenvalues associated with the reduced matrix $B^r$ obtained by eliminating the third, fourth, sixth, and seventh rows/columns from the matrix $B$, are $\{-40.1,  -10.0 ,  -5.50,   -4.40\}$, which ensures stability of the solution $\bar{\sigma}^r={B^r}^{-1} \bar{c}=[ 0.983,    0.274,    0.880,  0.880]$. Fig.\ \ref{A1b}(b) (\ref{A1b}(c)) shows the evolution of the individual $\mu_i(t)$ (of the payoffs $p_i(t)$). In contrast to the case of $b=1.1$, the nodes' payoffs converge on different final values: $p_1=1.13, p_2=0.652,p_3=0.411,p_4=0.411,p_5=1.21,p_6=1.29,p_7=1.29,p_8=1.21$. 
The different final payoffs of the players for $b=1.5$ are to be ascribed to their different \emph{locations} \footnote{Here, the \emph{location} of a node in the network is identified by the nodes it is connected to.} in the network. Note that we have repeated both the simulations in Figs.\ \ref{A1a} and \ref{A1b} many times, starting from different sets of initial conditions, and always observed convergence on a solution proportional to $\bar{v}_0$.

It is evident that fully mixed strategies (the fixed point observed in Fig.\ \ref{A1a}) and polarized strategies (the fixed point observed in Fig.\ \ref{A1b}) are substantially different. In particular, there are two main qualitative differences between the two,
\begin{enumerate}
  \item Fully mixed strategies are fixed points for which the payoffs are equal at all the network nodes, while polarized strategies are fixed points for which the payoffs are uneven.
  \item Fully mixed strategies are sensitive to the form of the network and the parameters of the game. Therefore it is possible to adjust/control a fully mixed strategy by changing the network connections and/or the parameters of the
game.  However, for polarized strategies, there are some players whose strategies cannot be changed, and thus the full state cannot be adjusted.
\end{enumerate}

Fully mixed strategy represent a fair situation, in which all the nodes achieve equal payoffs, regardless of their location in the network. On the other hand, polarized strategies represent an unfair situation, in which the nodes' payoffs vary according to their respective locations in the network. Therefore, it becomes important to understand how parameters of the game and variability of the networks yield stability of the fully mixed strategy.


The effects of varying the parameter $b$ (benefit of mutual cooperation) on stability of the fully mixed strategy are reported in Fig.\ 4.
Figure 4(a) shows the result of evolving the network in Fig.\ 2(d) for a long time from random initial conditions in the range $(0,1)$ as a function of the game parameter $b$ for $c=1$. {At the end of each run we record the vector of the final strategies $\bar{\sigma}_f$ and we plot the angle $A$
between the final state $\bar{\sigma}_f$ and the state associated with the eigenvector $\bar{v}_0$,
\begin{equation}\label{A}
A=\pi^{-1} \arccos\Big( \frac{\bar{\sigma}_f \cdot \bar{v}_0}{\|\bar{\sigma}_f \| \|\bar{v}_0 \|} \Big),
\end{equation}
where we have indicated with the symbol, $\cdot$, the dot product between vectors and with $\|\cdot\|$ the Euclidean norm of vectors.}
$A$ is close to zero for $b \lesssim 1.15$ and grows for larger $b$.  This is predicted by our stability criteria (see Fig.\ 4(b)). In Appendix C we show that in the limit in which $b=c$, the fully mixed strategy is stable. We expect this property to hold for $b$  \emph{close} to $c$. However, for $b$ above a critical value, the fully mixed state becomes unstable. This is shown in Fig.\ 4(d) where the largest nonzero eigenvalue of the matrix $B$ is plotted versus $b$.  The dashed section of the curve (starting at $b \simeq 1.15$) represents the range over which the entries of the eigenvector $\bar{v}_0$ are \emph{not} all of the same sign. {Thus we define \emph{the range of stability} of a fully mixed strategy, $R$, as the range in the game parameter $b$ (for $b>c$) such that the largest nonzero eigenvalue of $B$ is negative and the entries of the eigenvector $\bar{v}_0$ are of the same sign.}

An interesting result of Figs.\ 2, 3 and 4 is that by increasing the benefit to cost ratio $b/c$, the fully mixed state may be destabilized. For example, in Fig.\ 3 ($b=1.5$ larger than $b=1.1$, shown in Fig.\ 2), the fully mixed strategy is replaced by
a stable polarized strategy, for which some of the nodes converge on the pure cooperator state and some on the pure defector state.
Our interpretation of this phenomenon is the following. Fully mixed strategies are equilibria that arise in the presence of a balance between the benefit of being connected to a cooperator $b$ and the cost of cooperating $c$. The benefit to cost ratio $b/c$ affects stability of these equilibria, i.e., for a low ratio ($b/c$) they are stable and for a large ratio, they are unstable.

\begin{figure}[t]
\centering
\includegraphics[width=0.35\textwidth]{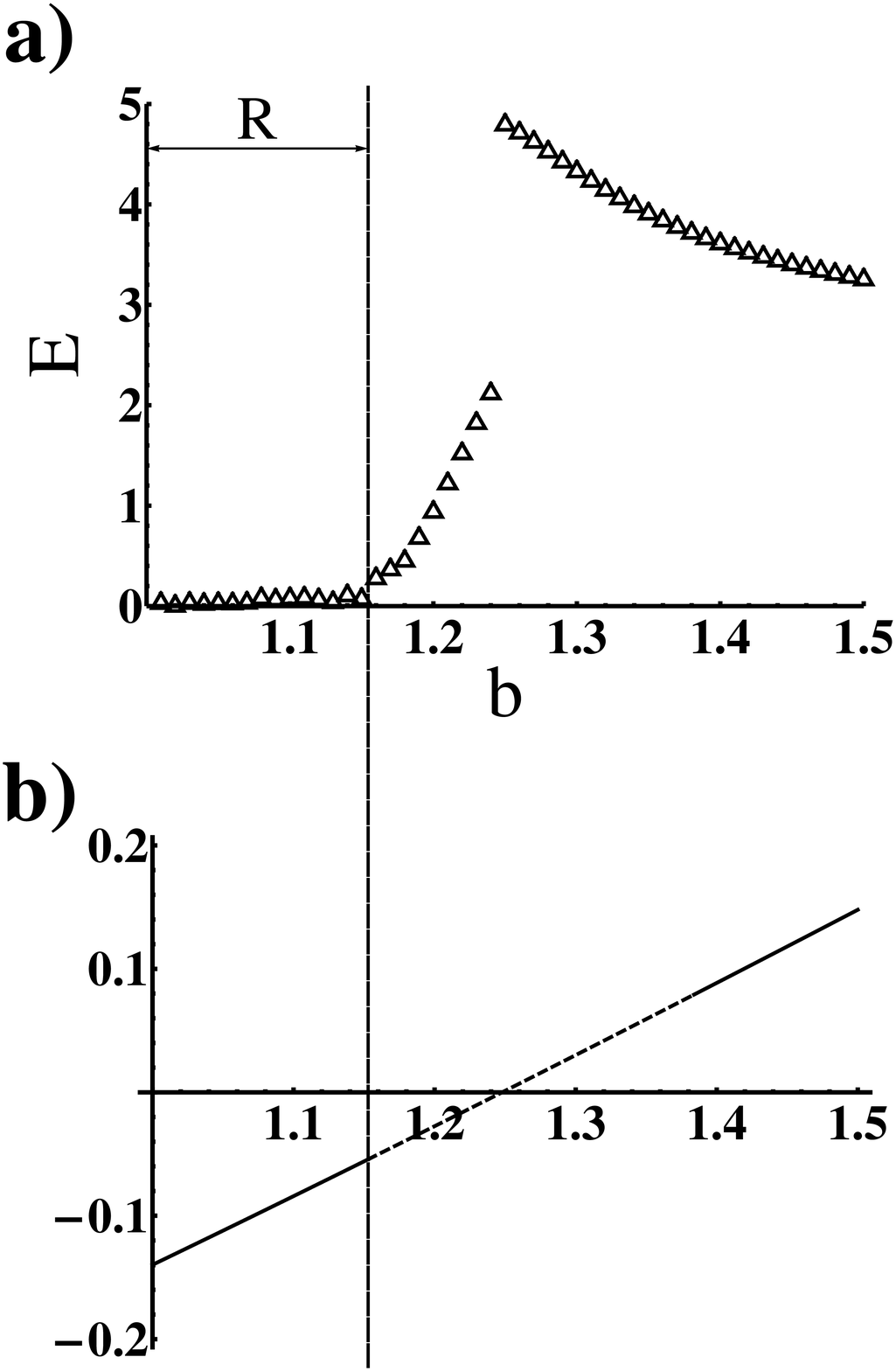}
\caption{\small  Plot (a) shows the result of evolving the network in Fig.\ 2d for a long time from random initial conditions in the range $(0,1)$. We plot $A$, the angle between the final state $\bar{\sigma}_f$ and the state associated with the eigenvector $\bar{v}_0$ as a function of the game parameter $b$ for $c=1$. The label $R$ represents the range of stability.  Plot (b) shows the largest nonzero eigenvalue of the matrix $B$ versus $b$.  The dashed section of the curve represents the range over which the entries of the eigenvector $\bar{v}_0$ are \emph{not} all of the same sign. }
\label{A2a}
\end{figure}

\section{Effects of the network topology on stability}


Fully mixed strategies are interesting equilibria, for which all the players' payoffs are equal. {Moreover, different from polarized strategies, fully mixed strategy states can be controlled by changing the network connections and/or of the parameters of the
game.} Thus, it is possible that according to the specific application of interest, these may be desirable/undesirable configurations.
In this section, we analyze how the underlying network structure can affect stability of fully mixed strategies. In particular, we analyze how the range of stability $R$, defined as the width of the $b$-range (see Fig.\ 4(a)) associated with stability of the fully mixed state, varies for different network topologies. We consider the case of symmetric network topologies, i.e., $A_{ij}=A_{ji}$, $i,j=1,...,N$, for which the eigenvalues $\{ \lambda_k \}$ of the matrix $B$ are real (see Appendix B).

We first analyze Erd\H{o}s-Renyi random graphs \cite{Er:Re}. These are networks for which any two nodes are connected by an edge with a constant probability $p$. Figure 5(a) shows the range of stability $R$ for Erd\H{o}s-Renyi random graphs of 200 nodes versus the probability of an edge parameter $p$. 
We see that the range of stability increases with $p$.
Notice that for networks with low connectivity $p$, it is unlikely to find a game for which a fully mixed strategy is stable.
In figure 5(b) we consider Erd\H{o}s-Renyi graphs of varying dimension $N$ and average degree fixed and equal $20$. The figure shows the range of stability $R$  versus $N$. 

We define the degree of node $i$ is defined as $k_i=\sum_j A_{ij}$. Heterogeneity in the degree distribution is probably the most
important feature that characterizes the structure of real networks.
The discovery that the basic structure of many real-world networks is
characterized by a power-law degree distribution, was pointed out by Barab{\'a}si and Albert in their seminal paper
\cite{Ba:Al99}, and has been verified by many observations of real networks. Specifically, the
analysis of data sets of biological, social and technological
networks has shown that these typically exhibit a
power-law degree distribution, $P(k) \sim k^{-\eta}$. Networks characterized by a power-law degree distribution are termed \emph{scale free}.

Plot 5(c) shows the range of stability $R$ for scale free networks of 200 nodes versus the exponent of the degree distribution, $\eta$, for which the average degree is kept fixed at 20.  The networks are generated by using the algorithm in Ref.\ \cite{korea}.  The range of stability increases with $\eta$.  This indicates that for networks with high heterogeneity (low $\eta$), we are unlikely to find a game (or we are unlikely to be playing a game) for which a fully mixed strategy is stable.

The form of the degree distribution
is an important property of the structure of a network. However, many other distinctive
properties have been uncovered to characterize the
structure of real networks in more detail, such as the formation of
communities of strongly interconnected nodes, frequently detected in
many real networks \cite{Ne:Gi02}, or particular forms of correlation or mixing among the network
vertices \cite{New03Mix}.

One measure of mixing is the correlation among pairs of linked nodes
according to some properties at the network nodes. A very simple
case is degree correlation \cite{New02Ass}, in which vertices choose
their neighbors according to their respective degrees. Nontrivial
forms of degree correlation have been experimentally detected in
many real-world networks, with social networks being typically
characterized by assortative mixing (which is the case where vertices
are more likely to connect to other vertices with approximately the
same degree) and technological and biological networks by
disassortative mixing (which takes place when connections are more
frequent between vertices of different degrees). In Ref. \cite{New02Ass}
degree correlation is measured by means of a single
normalized index, the Pearson statistic $r$ defined as follows:
\begin{equation}\label{r}
 r={1
\over{\sigma^2_q}}{\sum_{k,k'}kk'(e_{kk'}-q_k q_{k'})},
\end{equation}
where $q_k$ is the probability that a randomly chosen edge is
connected to a node having degree $k$; $\sigma_q$ is the standard
deviation of the distribution $q_k$ and $e_{kk'}$ represents the
probability that two vertices at the endpoints of a generic edge
have degrees $k$ and $k'$, respectively. Positive values of $r$
indicate assortative mixing, while negative values characterize
disassortative networks.

\begin{figure}[t]
\centering
\includegraphics[width=0.8\textwidth]{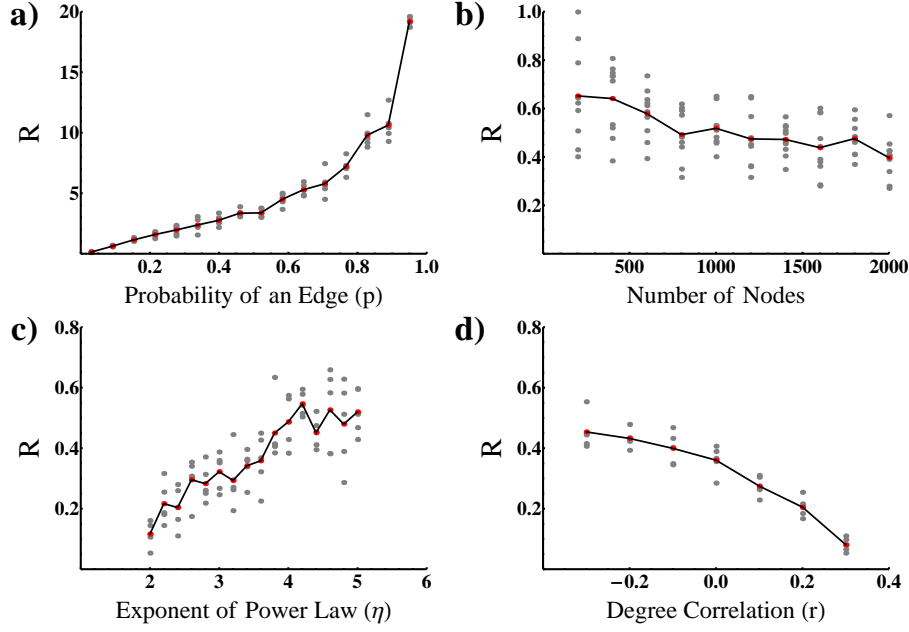}
\caption{\small   Plot (a) shows the range of stability $R$ (shown in Fig.\ 4(a)) for random Erd\H{o}s-Renyi graphs of 200 nodes versus the probability of an edge parameter $p$, for 5 different cases (grey points) for each $p$, with the average $R$ connected by line segments.  Plot (b) shows the range of stability $R$  for random Erd\H{o}s-Renyi graphs of dimension $N$ and average degree fixed and equal $20$, for 10 different cases (grey points) for each $N$, with the average $R$ connected by line segments.  Plot (c) shows the range of stability $R$  for scale free networks of 200 nodes versus the exponent of the degree distribution, $\eta$, where the average degree is kept fixed at 20.  Plot (d) shows the range of stability $R$ for scale free networks of 200 nodes,  $\eta=3.5$, average degree equal $20$, versus the degree correlation coefficient $r$.  }
\label{A2b}
\end{figure}


In Figure 5(d) we show the results of a numerical simulation in which we have kept the degree distribution fixed (power law with exponent $\eta=3.5$, average degree equal $20$) and we have made the coefficient $r$ vary from $-0.3$ to $0.3$ in steps of $0.1$ (for more details on the procedures that generates networks with different degree correlation properties, see Refs.\ \cite{New02Ass,reshuffling}). Specifically, we show that disassortative networks (i.e., $r$ negative) are characterized by a larger range of stability for the fully mixed strategy $R$ than their assortative counterparts.

\section{Conclusion}

In this paper, we have proposed a fully deterministic nonlinear model of an evolutionary game on a network, for which players are allowed  to pick  a strategy in the interval $[0,1]$, with $0$ corresponding to defection,  $1$ to cooperation, and intermediate values representing mixed strategies in which each player may act as a cooperator or a defector over a large number of interactions with a certain probability.
Our model is payoff-driven and it does not presume knowledge of the other players' strategies. Instead, we consider 
that strategies evolve based on a comparison of each player's payoff with those of his/her neighbors. We remove the unrealistic assumption that all the players are equal and assume that their choice of strategy depends on a parameter that takes into account the previous history of the game (in terms of payoffs). Under these assumptions, we find that fixed points of the dynamics may correspond to either one of two qualitatively different states: \emph{fully mixed strategies}, i.e.,  fixed points for which all the strategies are mixed and the payoffs are equal at different nodes, or \emph{polarized strategies}, i.e.,  fixed points for which at least one node is a full cooperator or defector and the payoffs are unequal at different nodes. We derive a simple condition that relates the network structure and the parameters of the game to stability of such fixed points, which provides a unified framework to study the effects of different network features, such as random, scale free, and degree-correlated topologies. This information can be used to predict which network/game combinations promote mixed versus polarized behavior.

In our simulations we choose a specific form for the function $f$, which is consistent with the assumption that cooperative behavior arises when the accumulated payoff exceeds a given threshold. Such a choice is supported e.g., by empirical studies that have found a positive correlation between the level of cooperation and the payoff for players participating in a Prisoner's Dilemma experiment \cite{SVO2}. Also, a natural interpretation of our proposed update mechanism is provided by Maslow's motivational theory \cite{Maslow} that assumes that  an individual's goals change according to whether certain basic needs are satisfied or not. In the language of game theory this corresponds to assuming that individuals start caring about the common good (cooperation) after they have consolidated their individual gain (defection).
However, our stability analysis is independent of the specific choice of the function $f$ and is valid over a broad range of possible functions $f$.


{We present a stability analysis for fixed points of Eqs. (\ref{d}).
We reduce the high-dimensional stability problem (\ref{delta}) in the low-dimensional form of Eq. (\ref{systemi}), which depends on the eigenvalues of a relevant matrix. These eigenvalues reflect the structure of the underlying network (i.e., the matrix $A$) and the choice of the parameters of the game (i.e., the two scalars $b$, and $c$). Similar reductions in a low-dimensional form have been proposed to evaluate (i) the stability of the synchronous evolution for networks of coupled oscillators \cite{FujiYama83,Pe:Ca,NSG,SAS,SOPO}, (ii) the stability of the consensus state in networks of coupled integrators \cite{CONS}, (iii) the stability of discrete state models of genetic control \cite{POM}, and (iv) the response of networks of coupled excitable systems to stochastic stimuli \cite{KI2}.}

Our computations provide evidence of the fact that for networks with a heterogeneous degree distribution (scale free) or networks with low connectivity, most games played will support the emergence of polarized strategies. However, for networks with a homogeneous degree distribution or networks with high connectivity, it is more likely that strategies will be mixed. We have also considered the effects of the network degree correlation and found that disassortative networks are characterized by a larger range of stability $R$ 
than assortative networks. Thus if we are given certain characteristics of a network (e.g., connectivity, the degree distribution, or the degree correlation), we may be able to determine whether polarized or mixed strategies will arise from the dynamics.  Alternatively, if we are given a game on a network, we may be able to modify the network to preferentially select polarized or mixed strategies depending on our application.

One surprising observation is that making the benefit $b$ larger than the cost $c$ may destabilize the mixed strategy state, with some players converging on the pure cooperator and some on the pure defector state. This suggests that policies governing dynamics on a network 
should consider the implications of making the benefit of cooperation too high.


The authors would like to thank Anurag Setty, Zeynep Tukekci, Gregory Taylor, Edward Ott, and Michelle Girvan for insightful discussions. {Both authors are indebted to
an anonymous reviewer for his valued comments and suggestions. }



\appendix
\section{The matrix B has at least one zero eigenvalue}

We define $(D-A)$ as the Laplacian matrix $L$.
We note that the sum of the elements of all the rows of the Laplacian matrix $L=(D-A)$ is equal zero. It follows that $L$ has at least one eigenvalue equal zero, with associated right eigenvector $v^r=[1,1,...,1]^T$ and left eigenvector $v^\ell$. Moreover, it can be shown that this zero eigenvalue is also the only one if the matrix $A$ is irreducible, i.e., the associated digraph is strongly connected. From $B^T=(bA-cD)^TL^T$, we see that $B^T {v^\ell}^T=(bA-cD)^T (L^T {v^\ell}^T)=0$ and therefore the matrix $B$ has at least one eigenvalue equal zero with associated left eigenvector $v^\ell$.

\section{A symmetric implies that B has real eigenvalues}

The property of the matrix $A$ of being symmetric does not imply that the matrix $B=(D-A)(bA-cD)$ is symmetric. However, it can be shown that if $A$ is symmetric, the spectrum of $B$ is real. In order to do this, we write the eigenvalue equation for the matrix $B$,
\begin{equation}
w^T (D-A)(bA-cD)=\lambda w^T, \label{ei}
\end{equation}
where $\lambda$ is a generic eigenvalue ($w^T$ the associated left eigenvector). Our goal is to show that $\lambda$ is real.
Recall that the graph Laplacian $L=(D-A)$ is symmetric and positive semidefinite \cite{WU_BOOK}. Hence, it can be decomposed as $Q Q^T$, where the matrix
$Q$ is also positive semidefinite . This yields,
\begin{equation}\label{sopra}
 w^T Q Q^T (bA-cD)   = \lambda  w^T.
\end{equation}
Right multiplying Eq. (\ref{sopra}) by $Q$, we obtain
\begin{equation}
u^T Q^T (bA-cD) Q  = \lambda u^T,
\end{equation}
where $u^T=w^T Q$. Since the matrix $Q^T (bA-cD) Q$ is symmetric it follows that $\lambda$ is real. 


\section{Stability conditions for $b=c$}

In the limit in which $b=c$, the matrix $B$ is equal to $B=-b(D-A)^2=-bL^2$. If the matrix $A$ is symmetric and irreducible, it follows that the eigenvalues of the matrix $B$ are negative and that the eigenvector associated with the only zero eigenvalue of the matrix $B$ is $[1,1,...,1]^T$. Thus the fixed point associated with this eigenvector is stable, as both conditions (II) and (IIIA) presented in the main manuscript, are satisfied.

\end{document}